\documentclass[conference, 10pt]{IEEEtran}

\usepackage{cite}
\usepackage{url}
\usepackage{graphicx}
\usepackage{color}
\usepackage{placeins}
\usepackage{float}
\usepackage{tabularx,colortbl}
\usepackage{ifthen}
\usepackage{amsmath}
\usepackage{amssymb}
\usepackage{float}
\usepackage{psfrag}

\usepackage{epstopdf}
\usepackage{pstool}

\usepackage{subcaption}
\usepackage{cite}

\makeatletter

\newcounter{author}
\renewcommand{\author}[2][]{
   \stepcounter{author}
   \@namedef{author@\theauthor}{#2}
   \@namedef{authorlabel@\theauthor}{#1}
}

\newcounter{address}
\newcommand{\address}[2][]{
   \stepcounter{address}
   \@namedef{address@\theaddress}{#2}
   \@namedef{addresslabel@\theaddress}{#1}
}

\newcommand{\alsep}{and}

\def\newmaketitle{\par%
  \begingroup%
  \normalfont%
  \def\thefootnote{}
  \def\footnotemark{}
  \let\@makefnmark\relax
  \footnotesize
  \footnotesep 0.7\baselineskip
  \normalsize%
  \twocolumn[\thenewmaketitle\@IEEEaftertitletext]%
  \if@IEEEusingpubid
     \enlargethispage{-\@IEEEpubidpullup}%
  \fi
  \endgroup
  \setcounter{footnote}{0}\let\maketitle\relax\let\@maketitle\relax
  \gdef\@thanks{}%
  \let\thanks\relax}

\def\thenewmaketitle{
  \newpage
  \begin{center}%
    \vskip0.2em{\Huge\@IEEEcompsoconly{\sffamily}\@IEEEcompsocconfonly{\normalfont\normalsize\vskip 2\@IEEEnormalsizeunitybaselineskip
   \bfseries\large}\@title\par}\vskip1.0em\par%
    \vspace{1ex}
    \newcounter{c@author}
    \newcounter{c@tmp}
    \ifthenelse{\value{author}=2}{%
      \newcommand{\liand}{ and }}{%
      \newcommand{\liand}{, and }}
    \ifthenelse{\value{address}<2}{%
      \@nameuse{author@1}%
      \stepcounter{c@author}%
      \whiledo{\value{c@author}<\value{author}}{%
        \setcounter{c@tmp}{\value{author}}%
        \addtocounter{c@tmp}{-\value{c@author}}%
        \ifthenelse{\value{c@tmp}=1}{%
          \renewcommand{\alsep}{\liand}}{\renewcommand{\alsep}{, }}%
        \stepcounter{c@author}\alsep \@nameuse{author@\thec@author}}\\%
    }
    {
      \@nameuse{author@1}${}^{(\ref{\@nameuse{authorlabel@1}})}$%
      \stepcounter{c@author}%
      \whiledo{\value{c@author}<\value{author}}{%
      \setcounter{c@tmp}{\value{author}}%
      \addtocounter{c@tmp}{-\value{c@author}}%
      \ifthenelse{\value{c@tmp}=1}{%
        \renewcommand{\alsep}{\liand}}{\renewcommand{\alsep}{, }}%
      \stepcounter{c@author}\alsep \@nameuse{author@\thec@author}%
        ${}^{(\ref{\@nameuse{authorlabel@\thec@author}})}$%
      }
    }
    \vspace{0.2ex}

    \ifthenelse{\value{address}>0}{%
      \ifthenelse{\value{address}=1}{
        {\@nameuse{address@1}}
      }
      {
        \newcounter{c@address}

        \begin{center}
        \whiledo{\value{c@address}<\value{address}}
        {
          \refstepcounter{c@address}
            ${}^{(\thec@address)}$\,%
              \label{\@nameuse{addresslabel@\thec@address}}%
              \@nameuse{address@\thec@address}\\ %
        }
        \end{center}
      } 
    }
    {
      \relax
    }
  \end{center}
}

\makeatother

\title{Bianisotropic Susceptibility Theory \\ of Artificial Magnetic Conductors}

\author[org1]{Oscar Céspedes Vicente}
\author[org1]{Christophe Caloz}

\address[org1]{Faculty of Engineering Science
Katholieke Universiteit (KU) Leuven
3001 Leuven, Belgium}

\begin{document}

\newmaketitle
\vspace*{1 cm}
\begin{abstract}
Over the past 30 years, Artificial Magnetic Conductors (AMCs) have generated immense interest in the microwave community. However, their description has been restricted to crude RLC design or subsidiary transmission-line network modelling. This paper introduces an \emph{electromagnetic} theory of AMCs in terms of bianisotropic susceptibilities.
\end{abstract}

\section{Introduction}
Due to the inexistence of magnetic charges, magnetic conductors, the dual of electric conductors, do not exist in a natural (atomic or molecular) form. However, artificial types of magnetic conductors, dubbed Artificial Magnetic Conductors (AMCs) and exhibiting quasi Perfect Magnetic Conductor (PMC) responses~\cite{kong1986electromagnetic,harrington2001timeharmonic} at specific frequencies, have been realized using resonant ``meta-particles'' of diverse shapes, including iris~\cite{walser1993new}, mushroom~\cite{sievenpiper1999high}, patch~\cite{zhang2003planar}, uniplanar cross potent~\cite{yang1999uniplanar} and grounded cross potent~\cite{hosseini2008characteristics}. Such AMCs have attracted immense attention due to diverse applications, such as low-profile antennas~\cite{feresidis2005artificial}, compact packaging~\cite{brazalez2012improved}, SAR reduction~\cite{raad2012flexible}, etc.

Unfortunately, the theory of AMCs has been so far mostly limited to crude RLC circuit modeling~\cite{sievenpiper1999high,yang1999uniplanar} or subsidiary transmission-line network modeling~\cite{rahman2001transmission,luukkonen2008simple}, and a more precise and fundamental theory -- namely a \emph{real electromagnetic theory}! -- has been seriously lacking.

This paper, leveraging the latest advances in metasurfaces~\cite{achouri2021electromagnetic}, provides such a theory by describing AMCs in terms of bianisotropic surface susceptibilities, with application to the mushroom AMC and a double-layer cross potent AMC. It paves the way for an even more complete electromagnetic theory that will include a precise explanation of generic AMC structures in terms of their dipolar responses.

\section{Selected AMC Structures}
\begin{figure}[H]
\vspace{1.5 cm}
\begin{center}
\noindent
\psfragfig[width=0.4\textwidth]{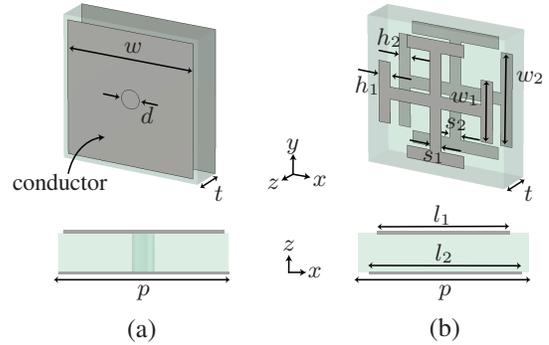}{
\psfrag{w}[c][c]{\small $w$}
\psfrag{d}[c][c]{\small $d$}
\psfrag{p}[c][c]{\small $p$}
\psfrag{t}[c][c]{\small $t$}
\psfrag{U}[c][c]{\small $w_{1}$}
\psfrag{I}[c][c]{\small $w_{2}$}
\psfrag{H}[c][c]{\small $s_{1}$}
\psfrag{J}[c][c]{\small $s_{2}$}
\psfrag{A}[c][c]{\small $h_{1}$}
\psfrag{S}[c][c]{\small $h_{2}$}
\psfrag{L}[c][c]{\small $l_{1}$}
\psfrag{K}[c][c]{\small $l_{2}$}
\psfrag{c}[c][c]{\small conductor}
\psfrag{a}[c][c]{(a)}
\psfrag{b}[c][c]{(b)}
\psfrag{x}[c][c]{\small $x$}
\psfrag{y}[c][c]{\small $y$}
\psfrag{z}[c][c]{\small $z$}
}         
\caption{Unit cell of representative AMC structures considered in the paper. (a)~Mushroom structure. (b)~Double-layer (asymmetric) cross potent structure.}
\label{fig:two_structures}
\end{center}
\end{figure}

The theory presented in this paper is completely general. It applies to any AMC structure. However, we shall consider the two particular structures shown in Fig.~\ref{fig:two_structures} as representative examples in order both to make the theory more concrete and provide study cases for the full-wave illustration of Sec.~\ref{sec:FW_ill}. The first structure, shown in Fig.~\ref{fig:two_structures}(a), is the mushroom (or Sievenpiper) AMC structure, while the second structure, shown in Fig.~\ref{fig:two_structures}(b), is a double-layer cross potent AMC structure.

\section{Susceptibility Analysis}
We assume, by duality with a conventional PEC, a PMC~\cite{kong1986electromagnetic,harrington2001timeharmonic} AMC that is both reciprocal and non-gyrotropic. Moreover, we restrict our attention, due to lack of space, to the case of s-polarization with scattering in the $zx$-plane (field components $E_y$, $H_z$ and $H_x$) and illumination under an angle~$\theta$ with respect to the normal of the structure. Finally, we write the ``metasurface'' incident, reflected and transmitted electric and magnetic field expressions and enforce the PMC boundary condition as $R=1$ and $T=0$, where $R$ and $T$ are the reflection and transmission coefficients, respectively, on the incident side of the structure.

Inserting the average and difference fields that incorporate all these conditions into generalized sheet transition conditions (GSTCs)~\cite{achouri2021electromagnetic} placed at in the top layer of the structure leads, after some algebraic manipulations, to the general AMC susceptibility condition
\begin{subequations}
    \begin{align}
		\chi_\text{ee}^{yy} &= \chi_\text{mm}^{zz} \sin^2(\theta), \\
		\chi_\text{me}^{xy} &=\frac{2j}{k}, \\
        \chi_\text{mm}^{xx}\in\mathbb{R}
        \quad&\text{and}\quad
        \chi_\text{em}^{yx}=-\chi_\text{me}^{xy}.
    \end{align}
 \label{PMC_susceptibilities}
\end{subequations}
\indent This condition contains two difficulties: 1)~it is undetermined, because it results from a system of two equations with four unknowns, and 2)~it is restricted to a specific angle, whereas we would ideally wish to obtain an angle-independent AMC response. Therefore, we shall adopt the alternative strategy consisting in first relaxing the AMC condition, next finding the general formulas for the angle-dependent reflection and transmission coefficients, and finally using the previous underdeterminacy to enforce the PMC condition at two separate angles. 

Using the same procedure as above, but with arbitrary $R$ and $T$ parameters and solving the resulting two equations for these two parameters yields
\begin{subequations}
	\begin{equation}
		\begin{split}
			R(\theta)=&-2\frac{k \chi_\text{ee}^{yy}+2 k \chi_\text{em}^{yx} c_\theta}{2 k \chi_\text{ee}^{yy}+2 k \chi_\text{mm}^{xx} c_\theta^2}\cdots\\
			&\ \frac{-k \chi_\text{mm}^{xx} c_\theta^2}{-j \left(k^2 \left(\chi_\text{ee}^{yy} \chi_\text{mm}^{xx}+(\chi_\text{em}^{yx})^2\right)-4\right)c_\theta}\cdots \\
			&\ \ \frac{-k_x \chi_\text{mm}^{zz} s_\theta}{+j k k_x \chi_\text{mm}^{xx} \chi_\text{mm}^{zz} s_\theta c_\theta-2 k_x \chi_\text{mm}^{zz} s_\theta}
		\end{split}
		\label{reflection_coeff}
	\end{equation}
        and\vspace{-5mm}
	\begin{equation}
		\begin{split}
			T(\theta)=&j\frac{4 c_\theta+k^2 \chi_\text{ee}^{yy} \chi_\text{mm}^{xx} c_\theta}{4 j c_\theta-j k^2 \chi_\text{ee}^{yy} \chi_\text{mm}^{xx} c_\theta}\cdots\\
			&\ \frac{k^2 (\chi_\text{em}^{yx})^2 c_\theta}{-j k^2 (\chi_\text{em}^{yx})^2 c_\theta+2 k \chi_\text{ee}^{yy}+2 k \chi_\text{mm}^{xx} c_\theta^2}\cdots\\
			&\ \ \frac{-k k_x \chi_\text{mm}^{xx} \chi_\text{mm}^{zz} s_\theta c_\theta}{-2 k_x \chi_\text{mm}^{zz} s_\theta+j k k_x \chi_\text{mm}^{xx} \chi_\text{mm}^{zz} s_\theta c_\theta},
		\end{split}
		\label{transmission_coeff}
	\end{equation}
\end{subequations}
where $c_\theta = \cos\theta$, $s_\theta = \sin\theta$ and $k_x=k\sin\theta$. Successively enforcing these two relations as $R=1$ and $T=0$ for both $\theta=0$ and $\theta=\theta_0$, where $\theta_0$ is an arbitrary angle in $]0,\pi/2]$, and solving the resulting system of (four) equations for the four susceptibilities, yields then
\begin{subequations}
	\begin{align}
		&\chi_\text{mm}^{xx} =0,\\
		&\chi_\text{ee}^{yy} =0,\\
		&\chi_\text{em}^{yx} =-\frac{2j}{k}=-\chi_\text{me}^{xy},\\
		&\chi_\text{mm}^{zz} =0 \label{eq:chi_mm}.
	\end{align}
 \label{AMC_susceptibilities}
\end{subequations}
These heteroanisotropic conditions are expected to represent, by design, conditions for an angle-independent PMC response.

\section{Full-Wave Illustration}\label{sec:FW_ill}

\begin{figure}[H]
     \begin{center}
\noindent
\psfragfig[width=0.38\textwidth]{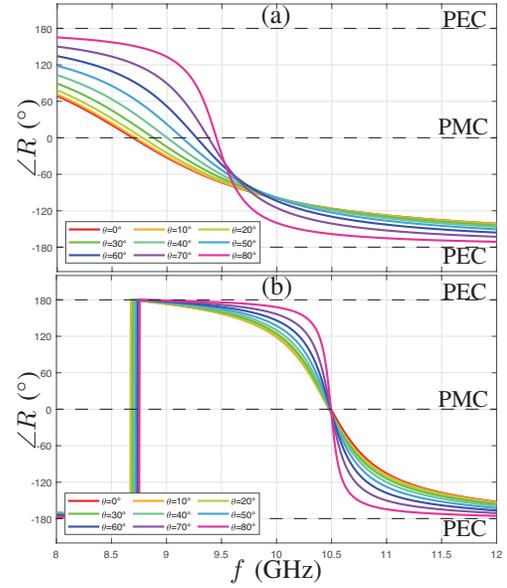}{
        \psfrag{a}[c][c]{(a)}
        \psfrag{b}[c][c]{(b)}
        \psfrag{R}[c][c]{$\angle R$ ($^\circ$)}
        \psfrag{f}[c][c]{$f$ (GHz)}
        \psfrag{E}[r][c]{\small PEC}
        \psfrag{M}[r][c]{\small PMC}
        }
        \caption{Phase of $R$ versus frequency for different illumination angles. (a)~Mushroom structure. (b)~Cross potent structure.}
    \label{fig:MushroomVsCrossPotent_Phase}
    \end{center}
\end{figure}

\noindent
Figure~\ref{fig:MushroomVsCrossPotent_Phase} plots \emph{optimized} AMC results for the two structures in Fig.~\ref{fig:two_structures}. The mushroom structure fails to exhibit an angle-independent response. The reason is that it fails to satisfy the condition~\eqref{eq:chi_mm}. In contrast, the double cross-potent structure does exhibit an angle-independent AMC response. This is because it satisfies the four conditions in~\eqref{AMC_susceptibilities}.
\\

\bibliographystyle{ieeetr}
\bibliography{AMC}{}
 
\end{document}